\documentclass[12pt]{iopart}

\usepackage{bm}
\usepackage{amssymb}
\usepackage{latexsym}
\usepackage{amsfonts}
\usepackage{epsfig}
\usepackage{psfrag}
\usepackage{xcolor}

\newcommand{\ket}[1]{\left|#1\right>}
\newcommand{\bra}[1]{\left<#1\right|}

\newcommand{\nn}{\nonumber\\}

\newcommand{\f}[1]{\mbox{\boldmath$#1$}}

\newcommand{\bea}{\begin{eqnarray}}
\newcommand{\eea}{\end{eqnarray}}

\newcommand{\abs}[1]{{\left| #1 \right|}}
\newcommand{\trace}[1]{{\rm Tr}\left\{ #1 \right\}}
\newcommand{\traceB}[1]{{\rm Tr_B}\left\{ #1 \right\}}
\newcommand{\ii}{{\rm i}}
\newcommand{\pdiff}[2]{\frac{\partial #1}{\partial #2}}

\begin{document}

\title{Relaxation Dynamics of Meso-Reservoirs}

\author{Gernot Schaller}
\ead{gernot.schaller@tu-berlin.de}
\author{Christian Nietner}
\author{Tobias Brandes}
\address{Institut f\"ur Theoretische Physik, Hardenbergstra{\ss}e 36,
Technische Universit\"at Berlin, D-10623 Berlin, Germany}
\pacs{67.85.-d, 51.10.+y, 05.60.Gg, 05.30.Rt}

\begin{abstract}
We study the phenomenology of maximum-entropy meso-reservoirs, where we assume that
their local thermal equilibrium state changes consistently with the heat transferred between the meso-reservoirs.
Depending on heat and matter carrying capacities, the chemical potentials and temperatures are
allowed to vary in time, and using global conservation relations we 
solve their evolution equations.
We compare two-terminal transport between bosonic and fermionic meso-reservoirs via systems that 
tightly couple energy and matter currents and systems that do not.
For bosonic reservoirs we observe the temporary formation of a Bose-Einstein condensate in one of the meso-reservoirs
from an initial nonequilibrium setup.
\end{abstract}

\maketitle

\section{Introduction}

Usually, a reservoir is treated as constant and inert to all systems that are coupled to it~\cite{Groot1984,nazarov2009quantum}.
In contrast, realistic experimental implementations always deal with finite-sized reservoirs~\cite{Fialko2012,Chien2012b,Donner2014}.
These are often too large to be treated exactly, but too small to neglect their dynamics 
in good faith, which has triggered research on thermodynamics with finite-size reservoirs~\cite{esposito2010c,Bruderer2012,reeb2014a}.
Especially noteworthy in this context are experimental setups which utilize ultra cold atoms embedded in optomagnetical traps or optical 
latices~\cite{Mandel2003,Seaman2007,Palzer2009,Brantut2012,Salger2013}.

In particular in nonequilibrium setups (e.g. realized by periodic driving or by locally different thermal states), 
also small systems may in the long-run transfer a significant amount of heat, 
and it may no longer be applicable to talk about a constant reservoir temperature or chemical 
potential~\cite{buettiker1993a,lim2013a,Brantut2013,Krinner2014,Donner2014,Grenier2014a}. 
For such reservoirs, we will use here the term meso-reservoir, with which we simply want to indicate that some sort of system back-action 
has to be taken into account, and that the state of a meso-reservoir is allowed to change in time. 
	
We assume that the system of interest can transfer entropy to the meso-reservoir in form of heat (in this paper, we will only consider matter and energy currents).
Moreover, we suppose that the meso-reservoir is subject to further processes that may potentially increase its entropy without additional heat transfer.
One possible microscopic example for such a process are interactions with a larger super-reservoir 
that leave energy and particle number invariant.
In~\ref{APP:puredephasing} we discuss this in detail for an energy-conserving interaction.
These processes generally lead in the eigenbasis of Hamiltonian $H$ and number 
operator $N$ to a fast decay of off-diagonal matrix elements 
(pure dephasing~\cite{skinner1986a,unruh1995a}), while the diagonals are by construction constant.
Since for any density matrix $\rho$, the entropy associated to the diagonal elements only, 
$S_D = - \sum_i \rho_{ii} \ln \rho_{ii}$ is larger than the entropy 
$S=- \trace{\rho \ln \rho}$ of the density matrix $S_D\ge S$~\cite{polkovnikov2011a}, pure 
dephasing processes will increase the entropy~\cite{morozov2012a} without injecting additional heat into the meso-reservoir.
Furthermore, it is well-known that almost all states appear thermal when sufficiently many degrees of freedom are traced out, 
a statement known as canonical typicality~\cite{Goldstein2006,Cramer2008,yukalov2011a,Reimann2012,Eisert2012,Pertot2014a}.
In usual derivations of master equations~\cite{Breuer} the reservoir is therefore always assumed in thermal equilibrium.
From the perspective of the meso-reservoir, the presence of the system will induce transfers between diagonal 
elements of the density matrix together with the exchange of matter and energy.
Additional elastic scattering processes within the meso-reservoir may support these equilibration processes.
It should be noted that this will always also generate off-diagonal matrix elements in the meso-reservoir
density matrix, see~\ref{APP:localrelaxation}.
We assume that these are quickly dephased.

In consequence, we do here as usual assume that the meso-reservoir is always kept at local equilibrium, i.e., 
at its maximum entropy state
\bea\label{EQ:maxent}
\rho_{MR} \propto e^{-\beta(t)[H_{MR} - \mu(t) N_{MR}]}\,,\qquad \trace{\rho_{MR}}=1\,,
\eea
where $\beta(t)=1/T(t)$ and $\mu(t)$ are now however time-dependent inverse temperature and chemical potential of the meso-reservoir, 
and $H_{MR}$ and $N_{MR}$ denote Hamiltonian and particle number operator of the meso-reservoir.
At this maximum entropy state, the internal entropy production of the meso-reservoir vanishes~\cite{Esposito2010a}, 
and the change of its entropy is solely governed by the heat transfer 
$\dot{S} \to \dot{S}_D = \beta(t)[J_E- \mu(t) J_M]$, quantified by the energy current $J_E$ and matter current $J_M$ entering the meso-reservoir via the system.
These energy and matter currents can be quantified for a large number of models~\cite{schaller2014}.
We note that energy contained in the interaction e.g. between system and meso-reservoir may in principle 
also affect its energy balance, but in the framework of our weak-coupling scenario we neglect these contributions in the long-term limit.

In this paper, we will consider the induced change of the meso-reservoir, which we compute self-consistently from the 
currents through the system. 
The system will only provide the dependence of the currents on temperatures and chemical potentials.
Therefore, we implicitly assume that the fastest timescale is the equilibration of the meso-reservoir to a thermal state~(\ref{EQ:maxent}).
Mainly for simplicity, we will also assume that the system quickly relaxes to its (possibly non-thermal) steady state, 
such that the current through the system has no signature of its initial state.
In this paper, we will be interested in the slow changes of the meso-reservoir parameters $\beta(t)$ and $\mu(t)$.

This paper is organized as follows.
In Sec.~\ref{SEC:equilib} we derive the differential equations determining the evolution of temperatures and chemical potentials in a general way.
In Sec.~\ref{SEC:fermionic_transport} we make these findings explicit for two fermionic meso-reservoirs coupled by single quantum dots.
In Sec.~\ref{SEC:bosonic_transport} we show how to treat bosonic meso-reservoirs including the possibility of Bose-Einstein condensation.
Finally, in Sec.~\ref{SEC:efficiency} we compare efficiencies of converting thermal gradients to chemical work.

\section{Consistent equilibrium states}\label{SEC:equilib}

Together with Eq.~(\ref{EQ:maxent}), the basic assumption of our framework is that particle number and energy content of 
meso-reservoirs $\alpha\in\{L,R\}$ (left, right) are given by integrals over densities of states 
${\cal D}_\alpha(\omega)$ and occupation numbers 
$n_\alpha(\omega)$ -- supplemented by a few states that are separately treated, e.g. the ground state
(thereby complementing previous work~\cite{gallego_marcos2014a})
\bea\label{EQ:particle_energy}
N_\alpha &=& \int {\cal D}_\alpha(\omega) n_\alpha(\omega) d\omega + n_\alpha^g\,,\nn
E_\alpha &=& \int {\cal D}_\alpha(\omega) \omega n_\alpha(\omega) d\omega\,.
\eea
Here, the density of states depends on dimensionality and character of the meso-reservoir -- but not on its
thermodynamic parameters $\mu_\alpha$ and $\beta_\alpha = 1/T_\alpha$ (we omit the explicit notion of time-dependence
for brevity).
In contrast, the occupation number depends explicitly on these
\bea
n_\alpha^\pm(\omega) = \frac{1}{e^{\beta_\alpha(\omega-\mu_\alpha)} \pm 1}\,,
\eea
where $n_\alpha^+$ corresponds to the Fermi-Dirac distribution in the fermionic and $n_\alpha^-$ to the Bose-Einstein distribution the bosonic case.
Furthermore, we note that for bosons we have $\mu_\alpha < 0$ and 
\bea
n_\alpha^g= \left[e^{-\beta_\alpha \mu_\alpha}-1\right]^{-1} = n_\alpha^-(0)
\eea
denotes the occupation of the ground state and thus allows the treatment of Bose-Einstein condensation~\cite{Anderson1995,Pethick}. 
For fermions, we set $n_\alpha^g=0$ (including it however would not substantially change the dynamics, since the Fermi-Dirac distribution is bounded). 
The change of particle numbers and energy content in every meso-reservoir has to balance the currents into each meso-reservoir, which gives -- when the currents are known -- 
rise to implicit ordinary differential equations for the thermodynamic potentials
\bea\label{EQ:odeimplicit}
\left(\begin{array}{c}
J_M^{(\alpha)}\\
J_E^{(\alpha)}
\end{array}\right)
= 
\left(\begin{array}{c}
\dot{N_\alpha}\\
\dot{E_\alpha}
\end{array}\right)
= C_\alpha
\left(\begin{array}{c}
\dot{\mu}_\alpha\\
\dot{T}_\alpha
\end{array}\right)\,,
\eea
where $C_\alpha$ is a $2\times 2$ capacity matrix for reservoir $\alpha$.
It can be split into a continuum contribution and a ground state contribution
\bea
C_\alpha = \int {\cal D}_\alpha(\omega) \left(\begin{array}{cc}
\pdiff{n_\alpha}{\mu_\alpha} & \pdiff{n_\alpha}{T_\alpha}\\
\omega \pdiff{n_\alpha}{\mu_\alpha} & \omega \pdiff{n_\alpha}{T_\alpha}
\end{array}\right) d\omega+\left(\begin{array}{cc}
\pdiff{n_\alpha^g}{\mu_\alpha} & \pdiff{n_\alpha^g}{T_\alpha}\\
0 & 0
\end{array}\right)\,.
\eea
A direct observation is that in contrast to classical quantities such as the geometric charge capacitance or heat capacity, 
the capacity matrix combines both temperatures and potentials to currents.
Its matrix elements will in general depend on temperatures and potentials, too, which e.g. is not the case for the 
geometric charge capacitance.
The integral term of the $11$-component is a continuum version of what is usually called quantum capacitance~\cite{iafrate1995a}.
It implicitly depends on the geometry via the chosen density of states, and is expected to approach the conventional
geometric capacitance in the limit of large $N_\alpha$.
In particular the separate treatment of the ground state in our case may however retain quantum features also in the macroscopic limit.

After fixing the density of states one can explicitly calculate the capacity matrix $C_\alpha$ -- for which
it is helpful to realize that the derivatives can be written as 
\mbox{$\pdiff{n_\alpha^\pm}{\mu_\alpha} = n_\alpha^\pm (1 \mp n_\alpha^\pm)\beta_\alpha$} and 
\mbox{$\pdiff{n_\alpha^\pm}{T_\alpha} = n_\alpha^\pm (1 \mp n_\alpha^\pm) \frac{\omega-\mu_\alpha}{T_\alpha^2}$} -- 
and the differential equation system~(\ref{EQ:odeimplicit}) can be made explicit by inverting $C_\alpha$.
Since the inverse of $C_\alpha$ also depends on $\mu_\alpha$, $T_\alpha$ and the specifics of ${\cal D}_\alpha(\omega)$, one thereby
obtains a coupled set of highly nonlinear differential equations.

When one now considers two meso-reservoirs $\alpha \in \{L,R\}$ coupled indirectly via the same system, 
such a two-terminal transport setup will obey conservation of total energy and matter
\bea
J_M^{(R)} &=& -J_M^{(L)} = J_M\,,\nn
J_E^{(R)} &=& -J_E^{(L)} = J_E\,,
\eea
which will lead to two conservation laws.
These imply that we can, in principle, eliminate two of the four thermodynamic variables to obtain two coupled nonlinear differential 
equations for e.g. the potential differences $V=\mu_L-\mu_R$ and temperature differences $\Delta T = T_L-T_R$.
Since the conserved quantities may be quite complex, we have however technically found 
it more convenient to evolve all four variables according to 
\bea
\left(\begin{array}{c}
\dot \mu_L\\
\dot T_L
\end{array}\right) 
= - C_L^{-1}
\left(\begin{array}{c}
J_M\\
J_E
\end{array}\right)\,,\qquad
\left(\begin{array}{c}
\dot \mu_R\\
\dot T_R
\end{array}\right) 
= + C_R^{-1}
\left(\begin{array}{c}
J_M\\
J_E
\end{array}\right)
\eea
and use the conservation laws as a numerical sanity check instead.
We see immediately that at configurations with vanishing currents the chemical potentials
and temperatures will remain stationary.
Normally, this can only be fulfilled at global equilibrium ($\mu_L=\mu_R$ and $T_L=T_R$), 
as the vanishing of both currents imposes two independent conditions.
However, in the tight-coupling regime ($J_E = \varepsilon J_M$), these conditions are not independent, and in consequence, stationary 
states may arise that are not global equilibrium states.

We note further that these equations could in principle be further simplified in the linear-response regime, 
where the currents are linear in potential and temperature differences~\cite{Onsager1931a,Nietner2013}.
However, also far away from this equilibrium regime, the currents must obey the second law (recall that $J_M$ and $J_E$ count positive when entering the right meso-reservoir), stating that the entropy production of the system
\bea\label{EQ:second_law}
\dot{S}_\ii = \left(\frac{1}{T_R(t)}-\frac{1}{T_L(t)}\right) J_E - \left(\frac{\mu_R(t)}{T_R(t)} - \frac{\mu_L(t)}{T_L(t)}\right) J_M \ge 0
\eea
is non-negative.

\begin{figure}[t]
\includegraphics[width=0.48\textwidth,clip=true]{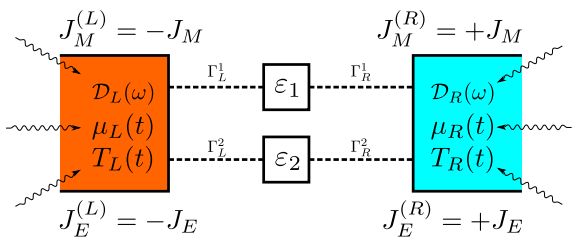}
\caption{\label{FIG:thermotransport_twoterminal}(Color online)
Sketch of the setups considered in this paper.
The meso-reservoirs are characterized by time-dependent temperatures $T_\alpha(t)$, 
time-dependent chemical potentials $\mu_\alpha(t)$, and the density of states ${\cal D}_\alpha(\omega)$.
The shown tunneling rates $\Gamma_\alpha^i$ enable for the exchange of energy and particles.
The individual currents through the two channels obey tight-coupling conditions, i.e., their matter and energy currents are proportional, 
but the combined current is not, $J_E \not\propto J_M$ (unless $\varepsilon_1=\varepsilon_2$).
Together with the entropy increase due to pure dephasing (wavy lines, see~\ref{APP:puredephasing}), these processes are assumed to lead to fast 
local equilibration of the meso-reservoirs.
}
\end{figure}

In the following, we will make this explicit for fermionic and bosonic reservoirs coupled 
via simple model systems, where we will assume that the energy and matter currents are in general
not tightly coupled $J_E \not\propto J_M$.
This is rather generic for realistic systems, but to keep the analysis simple, we consider coupling the meso-reservoirs via two non-interacting 
systems that -- when considered separately -- exhibit tight coupling~\cite{gomez_marin2006a}~(see Fig.~\ref{FIG:thermotransport_twoterminal}).
In conventional master equation derivations~\cite{Breuer}, our results can be obtained by performing the usual Born approximation for the 
full density matrix as $\rho(t) = \rho_S(t) \otimes \rho_{MR}(t)$ with Eq.~(\ref{EQ:maxent}), and for the 
Markov approximation assuming that $\rho_{MR}(t)$ changes even slower than the system density matrix $\rho_S(t)$, see~\ref{APP:masterequation}.
%


\section{Fermionic transport}\label{SEC:fermionic_transport}

For fermionic meso-reservoirs, we can relate the thermodynamic potentials with the currents using three standard integrals
\bea
J_M^{(\alpha)} &=& I_0^{(\alpha)} \frac{\dot{\mu}_\alpha}{T_\alpha} + I_1^{(\alpha)} \frac{\dot{T}_\alpha}{T_\alpha^2}\,,\\
J_E^{(\alpha)} &=& \left(I_1^{(\alpha)} + \mu_\alpha I_0^{(\alpha)}\right)\frac{\dot{\mu}_\alpha}{T_\alpha}
+ \left(I_2^{(\alpha)} + \mu_\alpha I_1^{(\alpha)}\right) \frac{\dot{T}_\alpha}{T_\alpha^2}\,,\nonumber
\eea
where 
\bea
I_n^{(\alpha)} &=& \int\limits_{-\infty}^{+\infty} {\cal D}_\alpha(\omega) (\omega-\mu_\alpha)^n n_\alpha^+(\omega)[1-n_\alpha^+(\omega)] d\omega\,.\;\;
\eea
In many solid state models one usually has only positive single particle energies (${\cal D}_\alpha(\omega<0)=0$), and the integrals can
be evaluated in this case too.
However, to illustrate the method we consider the simpler case of the complete wideband limit 
${\cal D}_\alpha(\omega) = D_\alpha$ (normally corresponding e.g. to a 2d free electron gas) also for negative frequencies.
Then, it is straightforward to show that the integrals become 
$I_0^{(\alpha)} = D_\alpha T_\alpha$, $I_1^{(\alpha)}=0$, and $I_2^{(\alpha)} = \frac{\pi^2}{2} D_\alpha T_\alpha^3$
(the same results would follow when only positive frequencies were allowed and the additional constraint $\mu_\alpha \gg k_{\rm B} T_\alpha$ was imposed).
Then, we obtain the capacity matrix
\bea
C_\alpha =
D_\alpha
\left(\begin{array}{cc}
1 & 0\\
\mu_\alpha & \frac{\pi^2}{3} T_\alpha
\end{array}\right)\,,
\eea
which can easily be inverted.
We see that in this particular case (negative energies or only positive energies with $\mu_\alpha \gg k_{\rm B} T_\alpha$) 
the $11$-component does not depend on temperatures or potentials -- just as the geometric capacitance.
Furthermore, the $22$-component is linear in the temperature, which is well-known for
the electronic heat capacity.
We also note that the capacity matrix becomes singular at zero temperature, the positivity of the system's 
entropy production~(\ref{EQ:second_law}) however ensures that the heat flow into a low temperature reservoir 
is always non-negative, and therefore the extreme zero-temperature limit cannot actually be reached.

Finally, we note that matter and energy conservation imply the conserved quantities
\bea
\bar{\mu} &=& \frac{D_L}{D_L+D_R} \mu_L + \frac{D_R}{D_L+D_R} \mu_R\,,\nn
E &=& \frac{D_L}{2} \mu_L^2 + \frac{D_R}{2} \mu_R^2+ \frac{D_L}{2} \frac{\pi^2}{3} T_L^2 + \frac{D_R}{2} \frac{\pi^2}{3} T_R^2\,.
\eea

\subsection{Quantum-Dot Coupling}

Our simplest example is the single-electron transistor in weak-coupling approximation.
Here, the current triggered by a single quantum dot hosting at most one electron is given by~\cite{schaller2014}
\bea
J_M = \gamma \left[n_L^+(\varepsilon)-n_R^+(\varepsilon)\right]\,,\qquad
J_E = \varepsilon J_M\,,
\eea
where the constant $\gamma$ depends on the details of the coupling between system and meso-reservoir
and $\varepsilon$ denotes the dot level.
These expressions also arise from the Landauer current formula~\cite{landauer1957a} when considering a strongly peaked transmission function.
Obviously, the currents will vanish when $\Delta T=0$ and $V=0$, but for the specific example
it is also possible to obtain a vanishing current whenever $T_R(\varepsilon-\mu_L) = T_L (\varepsilon-\mu_R)$.
The stationary state will therefore when plotted in the $V-\Delta T$-plane depend on the initial condition.

When we consider two quantum dots with on-site energies $\varepsilon_i$ that connect the meso-reservoirs in parallel but do not interact directly as
sketched in Fig.~\ref{FIG:thermotransport_twoterminal}, the individual currents just add
\bea
J_M &=& \sum_{i\in\{1,2\}} \gamma_i \left[n_L^+(\varepsilon_i)-n_R^+(\varepsilon_i)\right]\,,\nn
J_E &=& \sum_{i\in\{1,2\}} \varepsilon_i \gamma_i \left[n_L^+(\varepsilon_i)-n_R^+(\varepsilon_i)\right]\,,
\eea
and we see that the tight-coupling condition is not obeyed, i.e., $J_E \not\propto J_M$, when $\varepsilon_1 \neq \varepsilon_2$.
As before, the constants $\gamma_i=\Gamma_L^i \Gamma_R^i/(\Gamma_L^i+\Gamma_R^i)$ are given by the coupling details between
system and meso-reservoirs (compare Fig.~\ref{FIG:thermotransport_twoterminal}).
Since each quantum dot can host at most one electron, the currents remain finite at infinite external bias, 
(where $n_L^+(\varepsilon_i) \to +1$ and $n_R^+(\varepsilon_i) \to 0$).
In the loose coupling regime, the currents will in general only vanish when all thermodynamic parameters are equal, i.e., when $T_L=T_R$ and $\mu_L=\mu_R$.

\subsection{Meso-Reservoir Dynamics}

\begin{figure}[t]
\includegraphics[width=0.48\textwidth,clip=true]{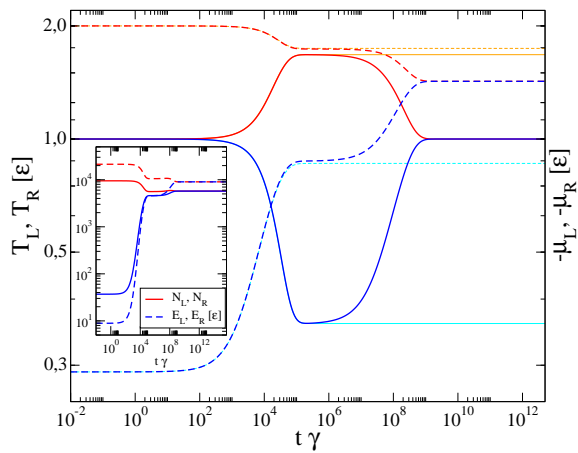}
\caption{\label{FIG:fermion_equilibration}(Color online)
Plot of time-dependent chemical potentials (solid) and temperatures (dashed) of the hot (red) and cold (blue) reservoirs for one fermionic 
transport channel with energy $\varepsilon_1 = \varepsilon$ (thin curves in lighter colors) and for two fermionic transport channels with energies 
$\varepsilon_1 = \varepsilon$ and $\varepsilon_2 = 1.1\varepsilon$ (thick curves).
At tight coupling (thin curves in lighter colors), a stationary nonequilibrium state is found.
Choosing the dot energies as different destroys the tight-coupling condition and leads to long-term 
equilibration (thick curves), but for intermediate times one observes a nonequilibrium pseudo-steady state where a potential bias is built up using the thermal gradient.
The inset shows the time evolution of the particle number (solid) and internal energy (dashed) of the reservoirs for two transport channels. 
Other parameters: $\gamma_1=\gamma_2 = \gamma/2$, $\mu_L^0=\mu_R^0=-\varepsilon$, $T_L^0=2\varepsilon$, $T_R^0=0.24\varepsilon$, 
capacity coefficients adjusted to $\varepsilon D_L= \varepsilon D_R=10000$ such that initially $N_L^0 \approx 9482$ and $N_R^0 \approx 37$.
}
\end{figure}

In Fig.~\ref{FIG:fermion_equilibration} we show the relaxation dynamics of the temperatures (dashed curves) and chemical potentials (solid curves) 
for two fermionic reservoirs in the wideband limit coupled via two non-interacting quantum dots.
Whereas for the tight-coupling configuration $\varepsilon_1=\varepsilon_2$ the stationary state of the complete system is a non-thermal nonequilibrium steady state 
(thin curves in lighter colors), 
the generic situation without tight-coupling (thick curves) yields an equilibrium state with $T_L=T_R$ and $\mu_L=\mu_R$.
However, for a near-tight-coupling configuration $\varepsilon_1 \approx \varepsilon_2$, one observes an intermediate pseudo-steady state before relaxation to complete equilibrium sets in at a much later time. 
The lifetime of the pseudo-steady state increases as one approaches the tight-coupling configuration, e.g., the smaller the difference $|\varepsilon_1 - \varepsilon_2|$ becomes.
During the evolution into this pseudo-steady state, the device has created a potential difference, thereby performing chemical work.
This is only possible with an initial temperature difference between the meso-reservoirs, and we will consider the
energetic efficiency of this process in Sec.~\ref{SEC:efficiency}.

\section{Bosonic transport}\label{SEC:bosonic_transport}

For bosons, we have to take into account some important differences:
First, the single-particle energies of the Hamiltonian must all be positive to bound the spectrum of the Hamiltonian.
Second, the chemical potentials must be negative to bound the occupation of the individual modes.
Finally, we allow for the possibility of a macroscopic occupation of the ground state $n_\alpha^g = \left[e^{-\beta_\alpha \mu_\alpha}-1\right]^{-1}$, 
which however does not significantly contribute to the total energy, cf. Eq.~(\ref{EQ:particle_energy}).
Then, we can relate the currents with the change of the thermodynamic parameters using just three integrals
\bea
J_M^{(\alpha)} &=& I_0^{(\alpha)} \frac{\dot{\mu}_\alpha}{T_\alpha} + (I_1^{(\alpha)} 
- \mu_\alpha I_0^{(\alpha)}) \frac{\dot{T}_\alpha}{T_\alpha^2}
+ \frac{e^{-\beta_\alpha \mu_\alpha}}{(e^{-\beta_\alpha \mu_\alpha}-1)^2} 
\left(\frac{\dot{\mu}_\alpha}{T_\alpha} - \frac{\mu_\alpha \dot{T}_\alpha}{T_\alpha^2}\right)\,,\nn
J_E^{(\alpha)} &=& I_1^{(\alpha)} \frac{\dot{\mu}_\alpha}{T_\alpha} + (I_2^{(\alpha)} - \mu_\alpha I_1^{(\alpha)}) \frac{\dot{T}_\alpha}{T_\alpha^2}\,,
\eea
where the integrals are given by
\bea
I_n^{(\alpha)} = \int_0^\infty {\cal D}_\alpha(\omega) \omega^n n_\alpha^-(\omega)[1+n_\alpha^-(\omega)] d\omega\,.
\eea
When the chemical potentials are negative one can obtain under additional assumptions on the density of states analytic expressions for these integrals.

We have also considered bosonic transport for a flat density of states (corresponding to 2d massive bosons, not shown), but here Bose-Einstein condensation will not occur.
Therefore, we consider an ohmic density of states instead.
With ${\cal D}_\alpha(\omega) = J_\alpha \omega$ (2d massless Bose gas supporting condensation~\cite{Pitaevskii}), the integrals become
\bea
I_0^{(\alpha)} &=& -J_\alpha T_\alpha^2 \ln\left(1-e^{\mu_\alpha/T_\alpha}\right)\,,\nn
I_1^{(\alpha)} &=& + 2 J_\alpha T_\alpha^3 {\rm Li}_2(e^{\mu_\alpha/T_\alpha})\,,\nn
I_2^{(\alpha)} &=& + 6 J_\alpha T_\alpha^4 {\rm Li}_3(e^{\mu_\alpha/T_\alpha})\,,
\eea
where ${\rm Li}_{n}(z) = \sum_{k=1}^\infty z^k/k^n$ denotes the polylog function~\cite{Abramowitz}.
In the high-temperature limit, the $22$-component of the capacity matrix simplifies to 
\mbox{$C_\alpha^{22} \to 6 J_\alpha \zeta(3) T_\alpha^2 + (- \mu_\alpha T_\alpha) J_\alpha \pi^2/3$}, with 
Riemann Zeta-function $\zeta(3)$ and has thus simple linear and quadratic contributions in the temperature.
For bosonic transport, the conserved quantities are given by
\bea
N &=& J_L T_L^2 {\rm Li}_2(e^{\mu_L/T_L}) + J_R T_R^2 {\rm Li}_2(e^{\mu_R/T_R})
+\frac{1}{e^{-\mu_L/T_L}-1} + \frac{1}{e^{-\mu_R/T_R}-1}\,,\nn
E &=& 2 J_L T_L^3 {\rm Li}_3(e^{\mu_L/T_L}) + 2 J_R T_R^3 {\rm Li}_3(e^{\mu_R/T_R})\,.
\eea
We define condensation by assuming that half of all meso-reservoir particles are in the ground state at negligible chemical potential, which
defines with ${\rm Li}_2(1) = \pi^2/6$ the condensation temperature
\bea
T_\alpha^{\rm cond} = \sqrt{\frac{3 N_\alpha}{J_\alpha \pi^2}}\,.
\eea
Since $N_\alpha(t)$ is a dynamic variable, this also transfers to the condensation temperature, such that one may also consider the condensate fraction (number of particles in the ground state of the reservoir versus total number of particles in each meso-reservoir) instead.

\subsection{Boson-Boson Transport model}

When the two bosonic meso-reservoirs are coupled via non-interacting harmonic oscillators, the
master equation currents can be written as
\bea
J_M &=& \sum_i \gamma_i \left[n_L^-(\varepsilon_i)-n_R^-(\varepsilon_i)\right]\,,\nn
J_E &=& \sum_i \varepsilon_i \gamma_i \left[n_L^-(\varepsilon_i)-n_R^-(\varepsilon_i)\right]\,,
\eea
which, similar to the fermionic case, can alternatively be obtained from the Landauer formula for heat transport~\cite{rego1998a,segal2003a}
in case of a strongly peaked transmission function.
In contrast to the fermionic case however, we observe that an infinite thermal bias 
(e.g. $n_L^-(\varepsilon_i) \to \infty$ and $n_R^-(\varepsilon_i) \to 0$) will let the currents diverge, too.
This essentially arises since the carrying capacity of the system between the reservoirs is not bounded.


\begin{figure}[t]
\includegraphics[width=0.48\textwidth,clip=true]{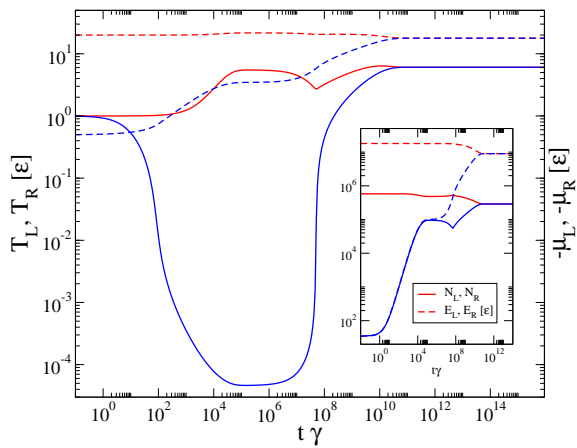}
\caption{\label{FIG:boson_equilibration}(Color online)
Plot of chemical potentials (solid) and temperatures (dashed) for the hot (red) and cold (blue) meso-reservoirs versus dimensionless time 
for two transport channels with energies $\varepsilon_1=\varepsilon$ and $\varepsilon_2=1.1\varepsilon$.
The inset shows the corresponding number of particles (solid) and the internal energy (dashed) of the reservoirs.
Other parameters: $\gamma_1=\gamma_2 = \gamma/2$, $\mu_L^0=\mu_R^0=-\varepsilon$, $T_L^0=20\varepsilon$, $T_R^0=0.5\varepsilon$, 
and capacity coefficients
$\varepsilon^2 J_L=\varepsilon^2 J_R=1000$, such that initially $N_L^0 \approx 577829$ and $N_R^0 \approx 35$.
}
\end{figure}

\subsection{Meso-Reservoir Dynamics}

Similar to the fermionic situation, the tight-coupling scenario will lead to a stationary
non-thermal steady state (not shown).
However, for slight modifications of the tight-coupling scenario, an
intermediate nonequilibrium state will emerge with a lifetime defined by the deviation from
tight coupling (see Fig.~\ref{FIG:boson_equilibration}).
Initially starting with a hot reservoir filled with many particles (red) and a cold reservoir with just a few particles (blue), 
one clearly observes that the initial thermal and particle gradients are used to dynamically induce condensation in the cold reservoir.
Eventually, the condensate evaporates again and global equilibrium is reached.

\begin{figure}[t]
\includegraphics[width=0.48\textwidth,clip=true]{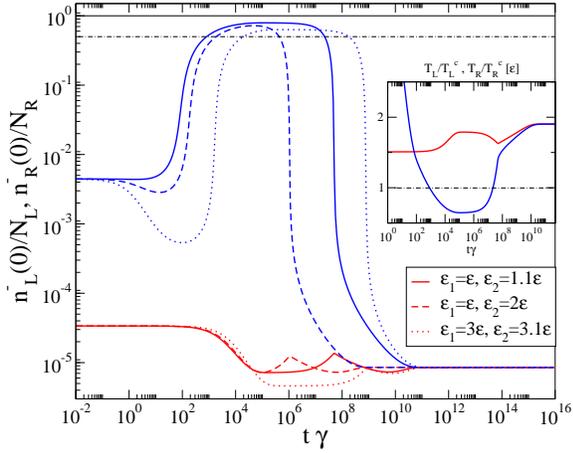}
\caption{\label{FIG:boson_groundstate}(Color online)
Plot of the fraction of particles which occupies the ground-state of the hot (red) and cold (blue) meso-reservoirs versus dimensionless 
time for different transport channel energies. 
The inset shows the time evolution of the reservoir temperatures normalized by their respective critical temperatures. We observe a temporarily decrease below a critical temperature leading to a macroscopic occupation of the ground-state energy level in the respective reservoir.
Other parameters: $\gamma_1=\gamma_2 = \gamma/2$, $\mu_L^0=\mu_R^0=-\varepsilon$, $T_L^0=20\varepsilon$, $T_R^0=0.5\varepsilon$, and 
capacity coefficients
$\varepsilon^2 J_L=\varepsilon^2 J_R=1000$, such that initially $N_L^0 \approx 577829$ and $N_R^0 \approx 35$.
}
\end{figure}

To evaluate the quality of the induced condensate, we have also investigated the condensate fraction 
for different transport channel configurations in Fig.~\ref{FIG:boson_groundstate}. 
For the case of a near tight-coupling configuration (solid) we observe a high quality condensate with about $80\%$ of the particles occupying the ground state.
This effect occurs due to the circumstance that the density in the cold reservoir grows faster than its temperature such that the condensation temperature is 
increased (inset) and Bose-Einstein condensation eventually sets in.
Further away from the tight-coupling configuration (dashed) the condensate quality as well as its lifetime is reduced.
Additionally, considering a near tight-coupling configuration with increased dot energies (dotted), we find that the condensate quality 
is further reduced, however it persists over longer times.

\begin{figure}[t]
\includegraphics[width=0.48\textwidth,clip=true]{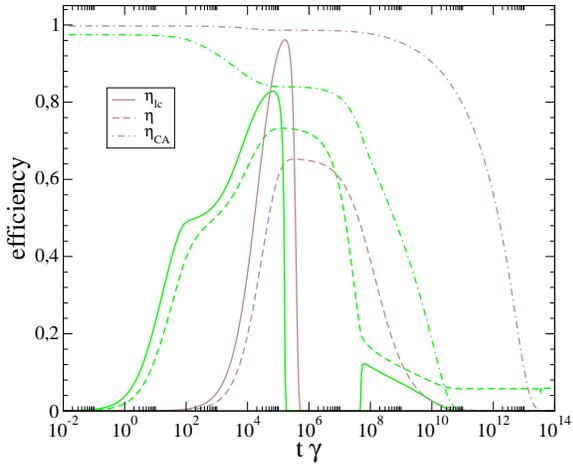}
\caption{\label{FIG:efficiency}(Color online)
Plot of the efficiency for bosonic (green) and fermionic (brown) meso-reservoirs with transport channel energies set 
to $\varepsilon_1=\varepsilon$ and $\varepsilon_2=1.1\varepsilon$. 
The time-local efficiencies (solid) are always upper-bounded by the time-local Carnot efficiencies (dash-dotted), 
and decay to zero for large times.
In contrast, the cumulative efficiencies (dashed) may remain finite.
We observe a temporarily negative chemical power output resulting from an inversion of the matter current direction.
For the fermionic reservoirs we set $T_L^0=84\varepsilon$, $T_R^0=0.24\varepsilon$, and capacity coefficients
$\varepsilon D_L= \varepsilon D_R=10000$, such that initially $N_L^0 \approx 577259$ and $N_R^0 \approx 37$.
For the bosonic reservoirs we set $T_L^0=20\varepsilon$, $T_R^0=0.5\varepsilon$, and capacity coefficients
$\varepsilon^2 J_L=\varepsilon^2 J_R=1000$, such that initially $N_L^0 \approx 577829$ and $N_R^0 \approx 35$.
Other parameters: $\gamma_1=\gamma_2 = \gamma/2$ and $\mu_L^0=\mu_R^0=-\varepsilon$.}
\end{figure}

\section{Efficiency}\label{SEC:efficiency}

In analogy to the intensively studied electronic solid-state setups, the transport setups suggested within this paper might be put to use as thermo-electric or
thermo-chemical generators~\cite{Seaman2007,Brantut2013,Heikkila2013}.
As a useful measure for the quality of such devices we consider the efficiency with which they generate power from an incoming heat current.

The internal energy of meso-reservoir $\alpha$ changes according to the fundamental equation (we have no volume change in the reservoirs) $dE_\alpha = T_\alpha dS_\alpha + \mu_\alpha dN_\alpha$.
Here, the term $T_\alpha dS_\alpha$ corresponds to the heat flow into the meso-reservoir, and
the term $\mu_\alpha dN_\alpha$ represents the chemical work~\cite{Esposito2009a}.
To define an energetic time-local efficiency, we consider the chemical power instead $P=J_M (\mu_R-\mu_L)$.
When the current flows from left to right $J_M>0$ although $\mu_R>\mu_L$, the power becomes positive, and the corresponding energetic efficiency is obtained by dividing by the heat flow entering the system from the hot (left) reservoir, i.e., for $T_L>T_R$ and $\mu_R > \mu_L$ one has for the efficiency~\cite{Esposito2009b}
\bea
\eta_{\rm lc}(t) = \frac{J_M(t) [\mu_R(t)-\mu_L(t)]}{J_E(t)-\mu_L(t) J_M(t)} \le \eta_{\rm CA}(t)\,,
\eea
where the bound by the time-dependent Carnot efficiency 
\bea
\eta_{\rm CA}(t) \equiv 1 - \frac{T_R(t)}{T_L(t)}
\eea
follows from the second law~(\ref{EQ:second_law}).
It is actually only reached in the tight-coupling case~\cite{gomez_marin2006a,schaller2014} (not shown).

In contrast, when one considers the cumulative efficiency, defined as ratio of total chemical work performed and
total heat influx from the hot (left) meso-reservoir up to time $t$ 
\bea
\eta(t) = \frac{\int_0^t J_M(t') [\mu_R(t')-\mu_L(t')] dt'}{\int_0^t [J_E(t')-\mu_L(t') J_M(t')] dt'}\,,
\eea
it follows directly from $J_M(t) [\mu_R(t)-\mu_L(t)] \le [J_E(t)-\mu_L(t) J_M(t)] \eta_{\rm CA}(t)$ 
that a (weak) upper bound is given by the initial Carnot efficiency
$\eta(t) \le \eta_{\rm CA}(0)$.

In Fig.~\ref{FIG:efficiency}, we show the resulting efficiencies for a fermionic setup in the wideband limit (brown) and 
for a bosonic setup with an ohmic density of states (green). 
In both cases, we observe an increase of the time-local efficiencies (solid) with time, getting ever 
closer to the respective Carnot efficiencies (dotted-dashed). 
However, at some specific time (around $t \gamma\approx10^6$) the power output becomes negative and, hence, the time-local efficiencies vanish. 
Only for the bosonic setup this behavior is reversed for even later times, leading to a finite time-local efficiency again.
Contrary, the cumulative efficiencies (dashed) are finite over a rather large time interval, and, 
moreover, they can have finite values even for arbitrary long times as can be seen for the bosonic setup. 

\section{Summary}

We have demonstrated that with a simple phenomenological approach conservation laws 
may be used to track the dynamical evolution of thermodynamic parameters of meso-reservoirs.
Our approach is applicable to a rather wide range of models, although we have exemplified it
only for two-terminal fermionic and bosonic transport setups and although we have for simplicity
neglected the energy and particle content of the system.
The generalization to other systems is straightforward, it is however necessary that the 
currents through the system obey the first law (conservation of matter and energy currents) and
the second law (to prevent unphysical temperatures and chemical potentials).
Naturally, when considering reservoirs of infinite capacities (formally e.g. by considering
the limit $D_\alpha,J_\alpha\to\infty$), temperatures and chemical potentials remain fixed and we
recover the usual weak-coupling master equation results.

In general, we have observed global equilibration of both meso-reservoirs in the long-term limit, 
except for the highly idealized tight-coupling scenario, where the non-equilibrium stationary state 
is frozen due to vanishing currents.
For situations close to tight-coupling, the system assumes a temporary pseudo-steady-state, 
and the dwell time of the system in this nonequilibrium state roughly depends on the deviation from 
the tight-coupling scenario.
We note that the dynamical generation of such nonequilibrium pseudo-steady state may be desirable
in many experimental contexts, and we have sketched the efficient use of such phases as a
thermo-electric generator and for preparing a Bose-Einstein condensate.

\section{Acknowledgements}

Financial support by the DFG (GRK 1558, SFB 910, SCHA 1646/2-1) is gratefully acknowledged.
The authors have also profited from discussions with F. Gallego-Marcos.


\section{Bibliography}

\bibliographystyle{iopart-num}
\bibliography{references}

\appendix
\section{Pure dephasing dynamics}\label{APP:puredephasing}

By pure-dephasing interactions for bipartite systems (denoted by $A$ and $B$)
we consider models where the interaction Hamiltonian $H_{AB}$ commutes with the Hamiltonian $H_A$, i.e., 
$[H_A, H_{AB}]=0$.
For simplicity, we do not consider the exchange of particles here (but the argument can be generalized).
With assuming an initially factorizing density matrix $\rho_0 = \rho_A^0 \otimes \rho_B^0$, the total solution
for the reduced density matrix of $A$ is then given by
\bea
\rho_A(t) = e^{-\ii H_A t} {\rm Tr_B}\left\{e^{-\ii (H_{AB}+H_B) t} \rho_A^0 \otimes \rho_B^0 e^{+\ii (H_{AB}+H_B) t} \right\} e^{+\ii H_A t}\,,
\eea
where we have also used that by construction $[H_A, H_B]=0$.
We now use that any interaction Hamiltonian can be written as
\bea
H_{AB} = \sum_\alpha A_\alpha \otimes B_\alpha\,,
\eea
where $A_\alpha=A_\alpha^\dagger$ act exclusively in the Hilbert space of $A$ 
and $B_\alpha=B_\alpha^\dagger$  only in the space of $B$.
Furthermore, assuming that $[H_A, A_\alpha]=0$ and $[A_\alpha, A_\beta]=0$ (which defines pure dephasing) we use the fact that
there is an eigenbasis diagonalizing all these operators, i.e., 
$H_A \ket{i} = E_i \ket{i}$ and $A_\alpha \ket{i} = \lambda_\alpha^i \ket{i}$.
The real numbers $E_i$ are the energies of $A$ and $\lambda_\alpha^i$ are the eigenvalues of the 
coupling operators.
Evaluating the density matrix of $A$ in this eigenbasis we obtain with $\bra{i} \rho_A \ket{j} = \rho_A^{ij}$
\bea
\rho_A^{ij}(t) = e^{-\ii (E_i-E_j) t} {\rm Tr_B}\left\{U_i(t) \rho_B^0 U_j^\dagger(t)\right\} \rho_A^{ij}(0)\,,
\eea
where the unitary operators are given by
\bea
U_i(t) = e^{-\ii(\sum_\alpha \lambda_\alpha^i B_\alpha + H_B) t}\,.
\eea
Clearly, one can see that the diagonal elements are just constant under the pure-dephasing assumptions, and hence
the energy of $A$ is not changed.
The matrix elements of unitary operators do have magnitude smaller than unity, from which one obtains that
$\abs{\trace{U_i \rho U_j^\dagger}} \le 1$ for any density matrix $\rho$.
Consequently, the absolute value of off-diagonal matrix elements can only decrease with respect to the initial state,
which a posteriori justifies the name pure-dephasing.
The situation we have in mind here is that of a meso-reservoir assuming the role of $A$ and an additional large 
reservoir $B$, whereby the recurrence time is sent to infinity and the reduction of off-diagonal matrix elements is very strong.
Indeed, one finds for specific models that the off-diagonal matrix elements simply decay exponentially~\cite{lidar2001a}.
Finally, we note that this statement does not rely on perturbative treatment and is thus valid beyond a master equation approaches.

\section{A local view on relaxation}\label{APP:localrelaxation}

In this section, we consider the possible transitions between matrix elements of a reduced density matrix.
With the same conventions as used in~\ref{APP:puredephasing}, the matrix elements of the reduced density
matrix evolve according to
\bea
\dot{\rho}_A^{ij} = -\ii (E_i-E_j) \rho_A^{ij} -\ii \bra{i} {\rm Tr_B}\left\{\left[H_{AB},\rho\right]\right\}\ket{j}\,,
\eea
where $\rho$ denotes the full density matrix.
We choose to represent its most general form by energy eigenstates of $A$
\bea
\rho = \sum_{ij} \rho_A^{ij} \ket{i} \bra{j} \otimes \rho_{(B,ij)}
\eea
where we use the convention that $\trace{\rho_{(B,ij)}}=1$ for all $i$ and $j$, such that in the
energy eigenbasis of $A$ we have the representation $\rho_A = \sum_{ij} \rho_A^{ij} \ket{i} \bra{j}$.
Inserting this decomposition and also the decomposition of the interaction Hamiltonian (which
now need not commute with $H_A$), we obtain
\bea
{\rm Tr_B}\left\{\left[H_{AB},\rho\right]\right\} = \sum_{k\ell} \sum_\alpha \left[A_\alpha, \rho_A^{k\ell} \ket{k}\bra{\ell}\right] {\rm Tr_B}\left\{B_\alpha \rho_{(B,k\ell)}\right\}\,.
\eea
For the dynamics of the reduced density matrix elements this implies
\bea
\dot{\rho}_A^{ij} &=& -\ii (E_i-E_j) \rho_A^{ij} - \ii \sum_k \left(\sum_\alpha \bra{i} A_\alpha \ket{k} {\rm Tr_B}\left\{B_\alpha \rho_{(B,kj)} \right\}\right) \rho_A^{kj}\nn
&&+ \ii \sum_k \left(\sum_\alpha \bra{k} A_\alpha \ket{j} {\rm Tr_B}\left\{B_\alpha \rho_{(B,ik)}\right\} \right) \rho_A^{ik}\,.
\eea
We note that this equation is non-perturbative in the interaction.
While it is probably useless for practical calculations, one can see that there is no direct coupling between different
diagonal elements.
This implies that to transfer population between different
diagonal elements $\rho_A^{ii}$ and $\rho_A^{jj}$ one has to populate also off-diagonal elements $\rho_A^{ij}$ as an intermediate step, too.

The basic assumption behind our Eq.~(\ref{EQ:maxent}) is that -- with $A$ describing the meso-reservoir and $B$ taking the role of the system -- additional 
pure dephasing processes as described in~\ref{APP:puredephasing} quickly eliminate the off-diagonal matrix-elements
in meso-reservoirs after local equilibration has been reached.
In contrast to usual derivations of master equations, we do however not 
neglect the energy (and particles) injected into the meso-reservoir.

\section{Derivation of a master equation}\label{APP:masterequation}

Here, we will follow the usual derivation of a master equation in the weak-coupling limit for 
time-dependent chemical potentials and temperatures.
We will perform the derivation only for a single meso-reservoir $B$ -- onto which in absence of stationary 
transport a small system $A$ would have negligible effect -- and highlight the changes arising from its time-dependence.
In the interaction picture (defined by bold-written operators) $\f{A}(t) = e^{+\ii (H_A + H_B) t} A e^{-\ii (H_A + H_B) t}$, the complete
density matrix follows the von-Neumann equation $\dot{\f{\rho}} = -\ii [\f{H_{AB}}(t), \f{\rho}(t)]$.
Re-inserting the formal solution in the right-hand side, one obtains
\bea
\dot{\f{\rho}} = -\ii [\f{H_{AB}}(t), \rho_0] - \int_0^t \left[\f{H_{AB}}(t), \left[\f{H_{AB}}(t'), \f{\rho}(t')\right]\right] dt'\,.
\eea
We now insert the Born approximation with a time-dependent reservoir density matrix $\f{\rho}(t) = \f{\rho_A}(t) \otimes \f{\rho_B}(t)$
and trace out the reservoir degrees of freedom $\f{\rho_A}(t) = \traceB{\f{\rho}(t)}$.
We note that due to trace conservation we have $\traceB{\f{\dot{\rho}_B}}=0$.
Furthermore, we assume that $\traceB{\f{B_\alpha}\rho_B^0}=0$, which is fulfilled for many microscopic models from the start but
can always be achieved by a suitable transformation.
Then we can insert the decomposition of the interaction Hamiltonian $\f{H_{AB}}(t) = \sum_\alpha \f{A_\alpha}(t) \otimes \f{B_\alpha}(t)$
to obtain an integro-differential equation (non-Markovian master equation) for the system density matrix
\bea
\dot{\f{\rho_A}} &=& \sum_{\alpha\beta} \int_0^t \Big[
+\f{A_\alpha}(t) \f{\rho_A}(t') \f{A_\beta}(t') \traceB{\f{B_\alpha}(t) \f{\rho_B}(t') \f{B_\beta}(t')}\nn
&&+\f{A_\beta}(t') \f{\rho_A}(t') \f{A_\alpha}(t) \traceB{\f{B_\beta}(t') \f{\rho_B}(t') \f{B_\alpha}(t)}\nn
&&-\f{A_\alpha}(t) \f{A_\beta}(t') \f{\rho_A}(t') \traceB{\f{B_\alpha}(t) \f{B_\beta}(t') \f{\rho_B}(t')}\nn
&&-\f{\rho_A}(t') \f{A_\beta}(t') \f{A_\alpha}(t) \traceB{\f{\rho_B}(t') \f{B_\beta}(t') \f{B_\alpha}(t)}\Big] dt'\,.
\eea
Next, we use the invariance of the trace under cyclic permutations and introduce the reservoir correlation function
$C_{\alpha\beta}(\tau,t') = \traceB{\f{B_\alpha}(\tau) B_\beta \f{\rho_B}(t')}$.
This requires to make use of $[H_{\rm B}, \f{\rho_B}(t)]=0$, cf. Eq.~(\ref{EQ:maxent}).
After the substitution $\tau=t-t'$, the master equation becomes 
\bea
\dot{\f{\rho_A}} &=& \sum_{\alpha\beta} \int_0^t d\tau \Big[
+\left[\f{A_\beta}(t), \f{\rho_A}(t-\tau) \f{A_\alpha}(t-\tau)\right] C_{\alpha\beta}(-\tau, t-\tau)\nn
&&+\left[\f{A_\beta}(t-\tau) \f{\rho_A}(t-\tau), \f{A_\alpha}(t)\right] C_{\alpha\beta}(+\tau, t-\tau)\Big]\,.
\eea
We apply the Markov approximation by assuming that the reservoir correlation function decays with respect to
its first argument much faster than $\f{\rho_A}(t-\tau)$ changes.
In fact, one can for many microscopic models explicitly confirm that the correlation function has
a Dirac-$\delta$-function-type behavior near $\tau=0$.
Since $\f{\rho_B}(t-\tau)$ changes even slower (the time-dependence in the second argument of the correlation function 
only refers to the change in temperatures and chemical potentials), this allows to replace 
$\f{\rho_A}(t-\tau) \to \f{\rho_A}(t)$ and $C_{\alpha\beta}(\pm\tau, t-\tau)\to C_{\alpha\beta}(\pm\tau, t)$
under the integral and to extend its upper bound to infinity, yielding a Markovian master equation
\bea
\dot{\f{\rho_A}} &=& \sum_{\alpha\beta} \int_0^\infty d\tau \Big[
+\left[\f{A_\beta}(t), \f{\rho_A}(t) \f{A_\alpha}(t-\tau)\right] C_{\alpha\beta}(-\tau, t)\nn
&&+\left[\f{A_\beta}(t-\tau) \f{\rho_A}(t), \f{A_\alpha}(t)\right] C_{\alpha\beta}(+\tau, t)\Big]\,.
\eea
Finally, we represent the system coupling operators in terms of eigenvectors of the system Hamiltonian
$\f{A_\alpha}(t) = \sum_{ij} A_\alpha^{ij} e^{+\ii(E_i-E_j) t} \ket{i}\bra{j}$ and neglect for large times
all terms that oscillate in $t$ (secular approximation), i.e., 
$\exp[\ii(E_i-E_j+E_k-E_\ell)t] \to \delta_{E_j-E_i,E_k-E_\ell}$, which yields with $L_{ab} = \ket{a} \bra{b}=L_{ba}^\dagger$
\bea
\dot{\f{\rho_A}} &=& \sum_{\alpha\beta} \sum_{ijk\ell} \delta_{E_k-E_\ell,E_j-E_i} 
A_\alpha^{ij} A_\beta^{k\ell} \times\nn
&&\times\Big[
+\left[L_{k\ell}, \f{\rho_A}(t) L_{ij}\right] \int_{-\infty}^0 d\tau e^{+\ii(E_i-E_j)\tau} C_{\alpha\beta}(+\tau, t)\nn
&&+\left[L_{k\ell} \f{\rho_A}(t), L_{ij}\right] \int_0^\infty d\tau e^{+\ii(E_i-E_j)\tau} C_{\alpha\beta}(+\tau, t)\Big]\,.
\eea
To see that this master equation is of Lindblad type, we can insert the even 
$\gamma_{\alpha\beta}(\omega,t) = \int C_{\alpha\beta}(\tau,t) e^{+\ii\omega\tau} d\tau$ 
and odd $\sigma_{\alpha\beta}(\omega,t) = \int C_{\alpha\beta}(\tau,t) {\rm sgn}(\tau) e^{+\ii\omega\tau} d\tau$ 
Fourier transforms of the reservoir correlation functions with
respect to their first argument
with which we can replace the half-sided Fourier transforms to yield
\bea
\dot{\f{\rho_A}} &=& \sum_{\alpha\beta} \sum_{ijk\ell} \delta_{E_k-E_\ell,E_j-E_i} 
\left(A_\alpha^{ji}\right)^* A_\beta^{k\ell} \frac{1}{2} \times\nn
&&\times\Big[
+\left[L_{k\ell}, \f{\rho_A}(t) L_{ji}^\dagger\right] \left[\gamma_{\alpha\beta}(E_i-E_j,t)-\sigma_{\alpha\beta}(E_i-E_j,t)\right]\nn
&&+\left[L_{k\ell} \f{\rho_A}(t), L_{ji}^\dagger\right] \left[\gamma_{\alpha\beta}(E_i-E_j,t)+\sigma_{\alpha\beta}(E_i-E_j,t)\right]\Big]\nn
&=& \sum_{\alpha\beta} \sum_{ijk\ell} \delta_{E_k-E_\ell,E_j-E_i} 
\left(A_\alpha^{ji}\right)^* A_\beta^{k\ell} \frac{\sigma_{\alpha\beta}(E_i-E_j,t)}{2} \left[\f{\rho_A}(t), L_{ji}^\dagger L_{k\ell}\right]\nn
&&+\sum_{\alpha\beta} \sum_{ijk\ell} \delta_{E_k-E_\ell,E_j-E_i} \left(A_\alpha^{ji}\right)^* A_\beta^{k\ell} 
\gamma_{\alpha\beta}(E_i-E_j,t)\times\nn
&&\times\Big[
L_{k\ell} \f{\rho_A}(t) L_{ji}^\dagger - \frac{1}{2} L_{ji}^\dagger L_{k\ell} \f{\rho_A}(t) - \frac{1}{2} \f{\rho_A}(t) L_{ji}^\dagger L_{k\ell}
\Big]
\eea
This is exactly the same Lindblad master equation as one would have obtained when assuming a constant reservoir and afterwards
inserting the time-dependent reservoir parameters~\cite{schaller2014}.
The used approximations do not go beyond those normally used in the derivation of master equation, except that some back-action
onto the reservoir is taken into account.

\end{document}